\documentclass[prl, nofootinbib, floatfix, notitlepage, twocolumn]{revtex4-1}

\usepackage{amsmath,amsfonts,amssymb}
\usepackage{mathrsfs}
\usepackage{graphicx}
\usepackage[english]{babel} 
\usepackage{hyperref} 
\usepackage{color}

\usepackage{animate}

\usepackage[cal=boondoxo]{mathalfa}

\hypersetup{
    colorlinks=true,
    linkcolor=red,
    citecolor=blue,
} 

\newcommand{\gy}{g_{Y}}

\begin{document}

\title{Gravitational Wave -- Gauge Field Oscillations}
\author{R. R. Caldwell}
\author{C. Devulder}
\author{N. A. Maksimova}
\affiliation{Department of Physics and Astronomy, Dartmouth College, 6127 Wilder Laboratory, Hanover, New Hampshire 03755 USA}

\date{\today}

\begin{abstract}
Gravitational waves propagating through a stationary gauge field transform into gauge field waves and back again. When multiple families of flavor-space locked gauge fields are present, the gravitational and gauge field waves exhibit novel dynamics. At high frequencies, the system behaves like coupled oscillators in which the gravitational wave is the central pacemaker. Due to energy conservation and exchange among the oscillators, the wave amplitudes lie on a multidimensional sphere, reminiscent of neutrino flavor oscillations. This phenomenon has implications for cosmological scenarios based on flavor-space locked gauge fields. 
\end{abstract}
\maketitle

In a remarkable series of papers starting with the work of Gertsenshteyn \cite{Gertsenshteyn1962}, the authors showed that a gravitational wave propagating through a stationary magnetic field converts into an electromagnetic wave and back again \cite{Poznanin1969,Boccaletti1970,Zeldovich1974}. Now that gravitational waves have been directly detected \cite{Abbott:2016blz}, no doubt there will be searches for this effect \cite{Dolgov:2012be}. 

Here we consider the more general phenomenon of the conversion of a gravitational wave into a stationary gauge field, as may be present in the early stages of the Universe \cite{Maleknejad:2011jw,Adshead:2012kp,Alexander:2014uza}. In particular, we show that gravitational waves transform into tensor waves of a gauge field, disappearing and reappearing much like neutrino flavor oscillations. More complicated oscillation patterns are possible for multiple families of gauge fields. Quantization of these gravitational and gauge field tensor modes reveals a novel relationship between the energy and flavor eigenstates that may leave an imprint on a spectrum of primordial gravitational waves, or even suggest a new mechanism for the origin of a primordial spectrum.

We consider a gauge field under general relativity,
\begin{equation}
S = \int d^4x \, \sqrt{-g} \left( \frac{1}{2}M_P^2 R - \frac{1}{4}\vec F_{\mu\nu} \cdot \vec F^{\mu\nu} + {\cal L}_m\right)
\label{eqn:action}
\end{equation} 
with metric signature $-+++$, $M_P$ is the reduced Planck mass, and ${\cal L}_m$ represents any other fields that may be present. We take an SU(2) field
\begin{equation}
\vec F_{\mu\nu} = \partial_\mu \vec A_\nu - \partial_\nu \vec A_\mu - \gy \vec A_\mu \times \vec A_\nu
\end{equation}
where $g_Y$ is the Yang-Mills coupling and the vector notation indicates direction in the three-dimensional flavor space. It is essential for this effect that the gauge field have a vacuum expectation value (vev), the analog of a stationary electric or magnetic field. To match the symmetries of our cosmological spacetime, we build a homogeneous and isotropic configuration as recently considered in the context of cosmic acceleration for inflation \cite{Maleknejad:2011jw,Adshead:2012kp,Alexander:2014uza} or dark energy \cite{ArmendarizPicon:2004pm,Mehrabi:2015lfa,Rinaldi:2015iza}. We work in the timelike gauge, so that $\vec A_t=0$, and require that the remaining components, one per each group generator, be spatially independent in order to ensure homogeneity. Isotropy is then achieved by identifying the global symmetry of SU(2) with the rotational O(3) symmetry of Euclidean space. Hence, we write  $\vec A_\mu = \phi(\tau) \vec e_\mu$ where $\vec e_\mu$ is a set of three mutually orthogonal, spacelike basis vectors. That is, the vector fields for flavors 1, 2, 3 point along the x, y, z directions, and we call this configuration {\it flavor-space locked}. The equation of motion in an expanding spacetime with line element $ds^2 = a^2(\tau)(-d\tau^2 + d\vec x^2)$ is $\phi'' + 2 \gy^2 \phi^3=0$, which may be solved exactly in terms of elliptic Jacobi functions. The classical field amplitude simply oscillates. The flavor-space locked configuration under this model is stable, as shown through a full cosmological perturbation analysis in Refs.~\cite{Bielefeld:2014nza,Bielefeld:2015daa}. Hence, we find that the gauge field strength tensor $\vec F_{\mu\nu}$ has non-zero components where we expect to find an electric field. Due to the coupling $\gy$ there is also a magnetic field which, for each flavor, is coaligned with the electric field. This vev enables the gauge field to support transverse, traceless, synchronous tensor fluctuations which couple to gravitational waves.

We generalize this scenario and find a much richer variety of behavior by considering multiple SU(2) subgroups embedded in a larger SU(N) gauge group. To do so, we consider its Cartan subalgebra whose $N-1$ dimensional basis elements each can be used to form linearly independent SU(2) subgroups,  ${\cal N}$ of which do not share any generators, where ${\cal N}$ is the largest integer less than or equal to ${N}/{2}$. Hence for SU(3) there is only ${\cal N}=1$ such subgroup, but for SU(4) and SU(5) there are ${\cal N}=2$.

As an example, in the case of SU(4), we identify the ${\cal N}=2$ subgroups by considering the $4 \times 4$ matrix representation of its $15$ generators. It is straightforward to find matrix representations of the two SU(2) subgroups which do not share any generators and are what we informally call ``non-overlapping" in that the structure constants $f_{abc}$ are zero when $a$ is from the first subgroup and $b$ is from the second subgroup. We use these two SU(2) subgroups of SU(4) to build a pair of isotropic, homogeneous, flavor-space locked field configurations. This amounts to singling out the x, y, z axes ${\cal N}$ times using the same mapping between the generators and real space, so that our construction is equivalent to the direct product of ${\cal N}$ SU(2) subgroups. The equation of motion for the field amplitude of each SU(2) subgroup $\phi_n$ for $n = 1,\,2,\,...\,, {\cal N}$ is similarly $\phi_n'' + 2 \gy^2 \phi_n^3=0$ so that each subgroup is independent of the others, and stable. For simplicity, we assume the gauge field vev has the same amplitude in all subgroups. 

As a worked example, we consider a gravitational wave and ${\cal N}$ gauge field waves propagating in an expanding spacetime with line element $ds^2 = a^2(\tau)(-d\tau^2 + d\vec x^2)$. We follow the notation in Ref.~\cite{Dimastrogiovanni:2012ew,Namba:2013kia,Bielefeld:2014nza,Bielefeld:2015daa}, where the behavior of the coupled system for ${\cal N}=1$ was studied in the context of inflation and other cosmological scenarios. We consider weak distortions of the spacetime metric and the gauge field, $\delta g_{\mu\nu} = a^2 h_{\mu\nu}$ and $\delta \vec A_\mu \cdot \vec e_\nu = a y_{\mu\nu}$, where  $h_{\mu\nu},\, y_{\mu\nu}$ are transverse, traceless, and synchronous. Since the Yang-Mills coupling breaks the chiral symmetry of left- and right-circular polarizations \cite{Adshead:2013qp}, it is practical to express the gravitational and gauge field waves with amplitudes $h_p,\, y_{pn}$ in a chiral basis with $p=L,\,R$ for $n=1,\, 2,\, ...\,,\, {\cal N}$. A further change of variables, $h = H \sqrt{2} / {a M_P} $ and $y = Y/{\sqrt{2} a} $, puts the action into canonical form. The Lagrangian for gravitational and gauge field waves, propagating with Fourier wave number $k$ much greater than both the expansion rate and the gauge field time rate of change, is given by  
\begin{eqnarray}
{\cal L} &&= \frac{1}{2}H_R'^2 - \frac{1}{2}k^2 H_R^2 + \sum_{n=1}^{\cal N}\left[ \frac{1}{2}Y_{Rn}'^2 - \frac{1}{2}k^2 Y_{Rn}^2\right.\cr
&&\left.+ k \gy \phi Y_{Rn}^2 - \frac{2}{aM_P}H_R\left(k \gy\phi^2Y_{Rn} + \phi' Y_{Rn}'\right)\right].\,\,\,
\label{eqn:LHY}
\end{eqnarray}
The prime indicates the derivative with respect to conformal time, $\tau$. The equations for $H_L,\,Y_{Ln}$ are obtained by replacing $k \to -k$ or $\gy \to -\gy$. The chiral asymmetry can be removed by setting $\gy=0$, which corresponds to {\it flavor electrodynamics} since the theory then consists of three copies of Maxwell electrodynamics. 

At the most basic level, the Lagrangian above describes ${\cal N}+1$ coupled oscillators, apart from the dynamics of the background field and cosmic expansion. At high frequency, $H$ and each of $Y_n$ oscillate with frequency $k$. The gravitational wave couples to each gauge field wave; each gauge field wave couples only to the gravitational wave. 
 
The  gravitational wave -- gauge field oscillations are revealed by the rms amplitude of the waves ${\cal h}$ and ${\cal y}_n$ in the high frequency limit. We write $H = {\cal h} e^{-i k \tau}$ and $Y_n = {\cal y}_n e^{-i k \tau}$ and choose $k$ to be sufficiently large such that we can treat the coefficients $A_1 = \phi'/a M_P$, $A_2 = \gy \phi^2/a M_P$, $A_3 = \gy \phi/2$ as constants. The resulting leading order solutions to the equations of motion are
\begin{eqnarray}
{\cal h}_R =e^{i A_3 \tau} &&\left[  c_0 \cos\omega \tau  - \frac{1}{\omega}\sin\omega \tau \right. \cr
&& \left.  \times\Big(  [ A_1+i A_2] {\cal N} c_{\cal N} +i A_3 c_0 \Big) \right] \label{eqn:h} \\
{\cal y}_{Rn} = e^{i A_3 \tau}&& \biggl[ (c_n -  c_{\cal N} )e^{i A_3 \tau} +  c_{\cal N}\cos\omega\tau \biggr. \cr
&&\left. +\frac{1}{\omega}\sin\omega\tau \Big([ A_1- i A_2 ]c_0 + i A_3 c_{\cal N} \Big)\right]
\label{eqn:yn}
 \end{eqnarray}
where $\omega^2 = {\cal N}(A_1^2 + A_2^2) + A_3^2$, and the initial amplitudes are ${\cal h}(0)=c_0,\, {\cal y}_n(0)=c_n$, and $c_{\cal N} = {\cal N}^{-1}\sum_{n=1}^{\cal N} c_n$. Hence, ${\cal h}$ and ${\cal y}_n$ oscillate at rates set by the gauge field vev, such that ${\cal h}$ is out of phase with the ensemble of gauge fields ${\cal y}_n$. The magnitude of ${\cal h}$ oscillates with frequency $\omega$, whereas the gauge field amplitude oscillates with two frequencies, $\omega$ and $A_3$. 

The normal modes of the system of oscillators are identified by diagonalizing the Lagrangian (\ref{eqn:LHY}). Starting with the gravitational and gauge field modes $\psi^i = ({\cal h},\, {\cal y}_n)$ for $i=0,\,1, ...\,,\, {\cal N}$, we write the Lagrangian in the form ${\cal L} = \tfrac{1}{2}\psi'^\dagger \mathbb{I} \psi' - \tfrac{1}{2}\psi^\dagger M^2 \psi$. We transform into the eigenbasis of $M^2$ via $\psi^i = R^i_j \Delta^j$, where $\Delta^j= (\Delta_0,\, \Delta_n)$ are the normal modes. Hence, the Lagrangian acquires the form ${\cal L} = \tfrac{1}{2}\Delta'^\dagger \mathbb{I} \Delta' - \tfrac{1}{2}\Delta^\dagger \Omega^2 \Delta$ and $\Omega^2$ is diagonal. The normal mode frequencies are $\omega+A_3,\, \omega-A_3$, and $2 A_3$ with ${\cal N}-1$-fold degeneracy.  These modes are plainly seen to comprise the gravitational and gauge field solutions in Eqs.~(\ref{eqn:h}-\ref{eqn:yn}). 

\begin{figure}[h]
\center\includegraphics[width=0.45\textwidth,angle=0]{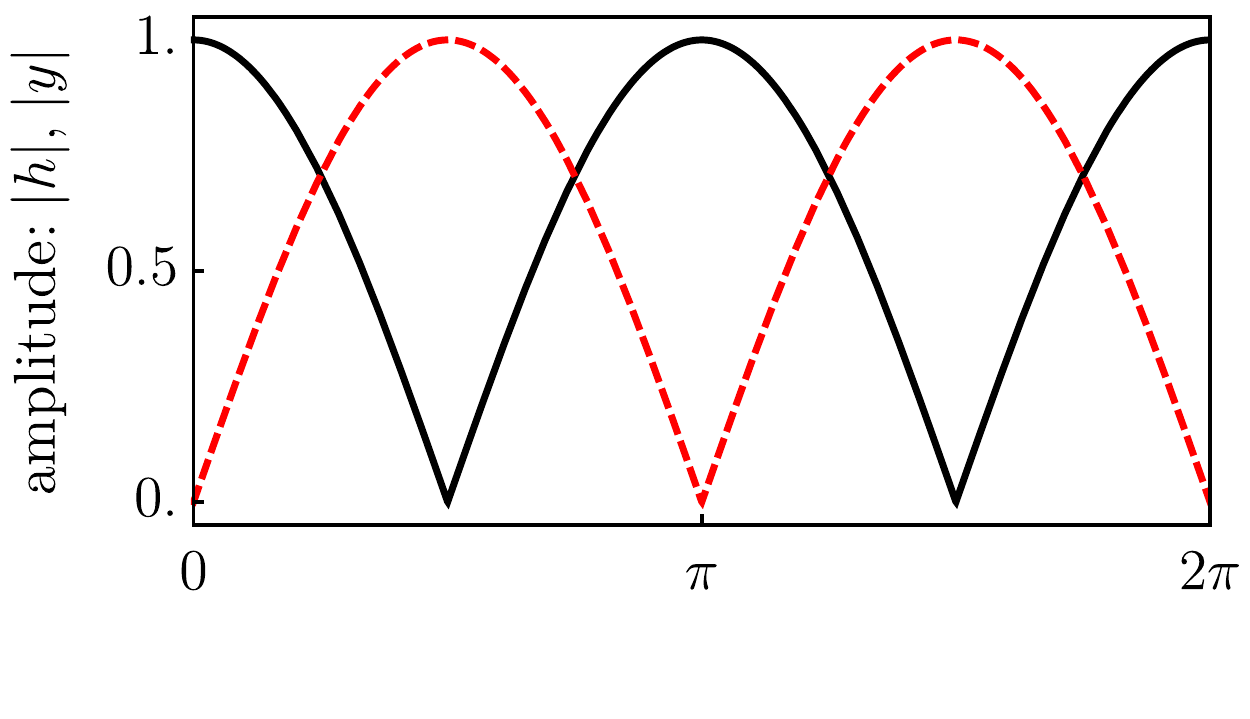}
\vspace{-0.75cm}
\center\includegraphics[width=0.45\textwidth,angle=0]{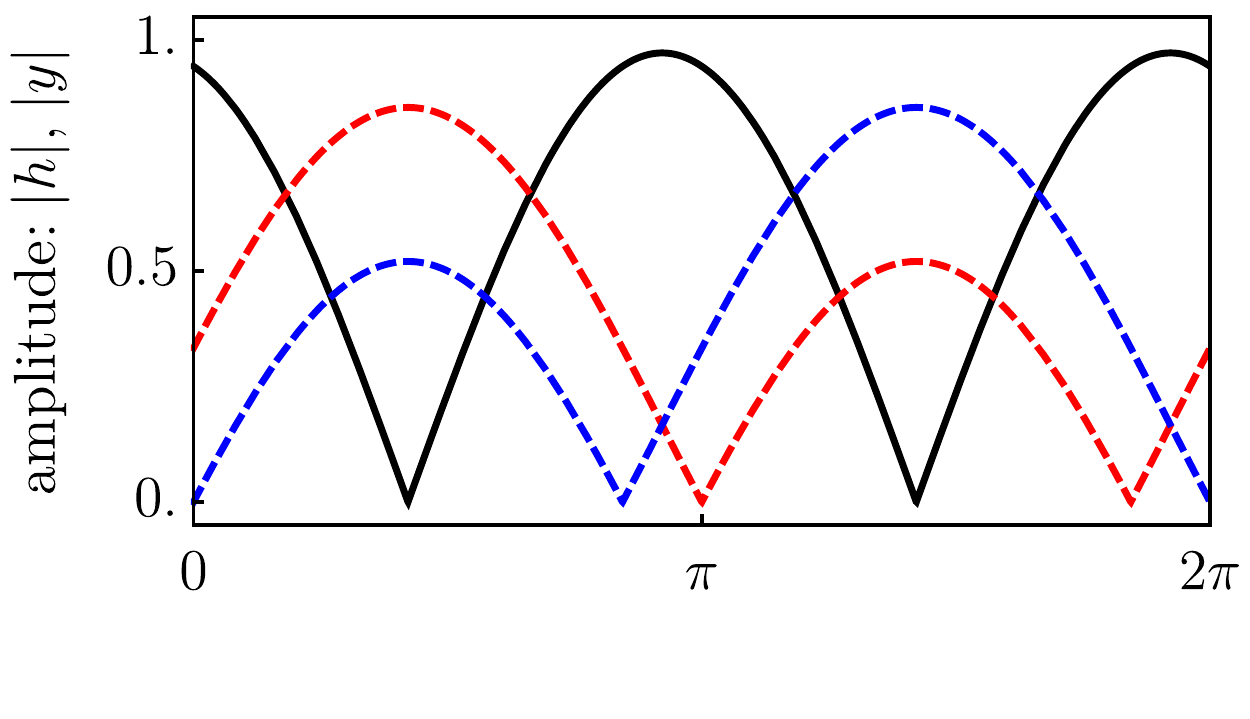}
\vspace{-0.75cm}
\center\includegraphics[width=0.45\textwidth,angle=0]{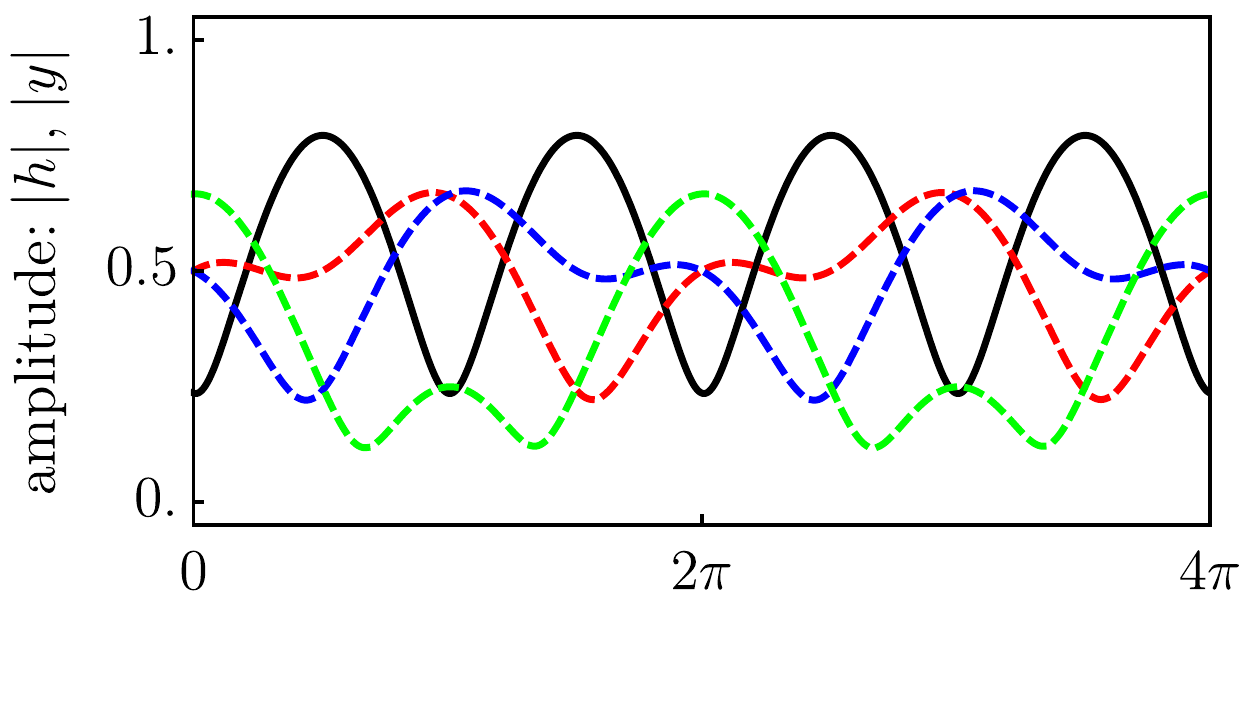}
\vspace{-0.75cm}
\center\includegraphics[width=0.45\textwidth,angle=0]{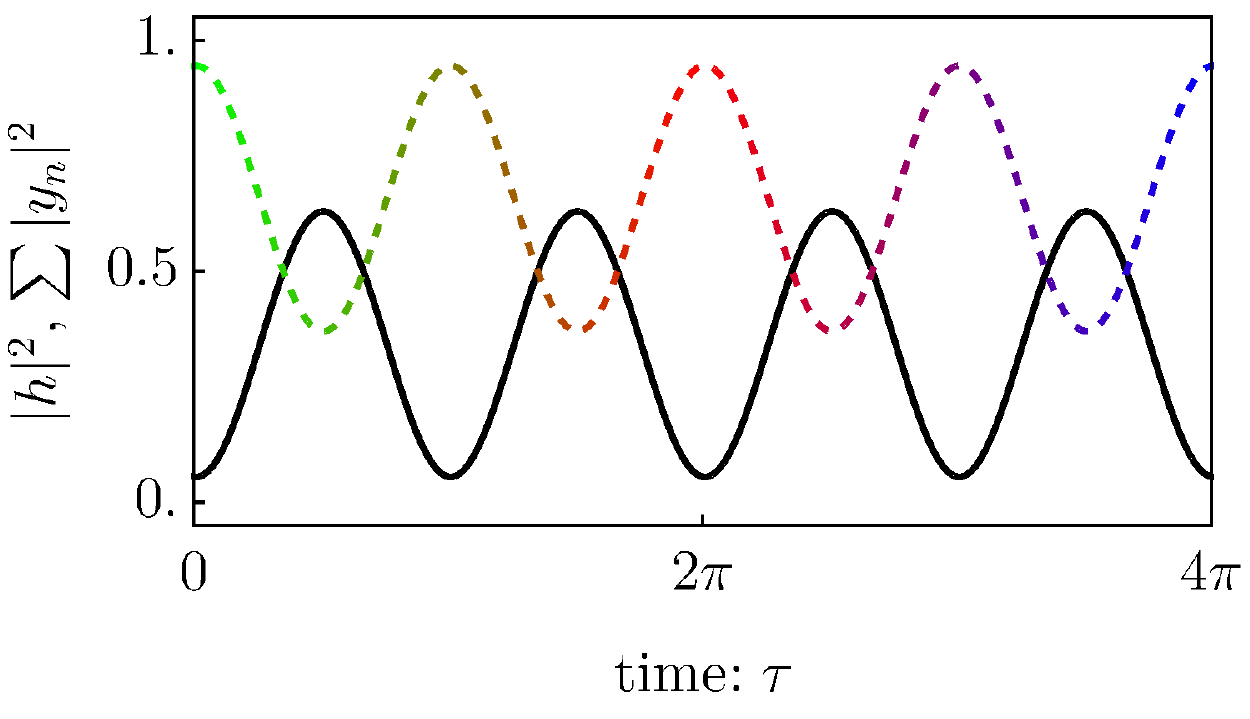}
\caption{The oscillations of the gravitational wave amplitude ${\cal h}$ (black) and gauge field ${\cal y}$ (dashed) are shown for ${\cal N}=1,\,2$ and $3$ (bottom two panels). In all cases, $|\psi|^2=1$ and the coefficients $A_i$ are scaled so that $\omega=1$. The bottom panel shows that the squared amplitudes sum to unity.}
\label{fig:fig1}
\end{figure}

Conservation of the canonical stress-energy tensor $\Theta^{\mu\nu} = \partial^\mu \psi^i \delta {\cal L}/\delta \partial_\nu \psi^i - \eta^{\mu\nu} {\cal L}$ yields the constant of motion in the high frequency limit, $|\psi|^2$. This conserved quantity is upheld by Eqs.~(\ref{eqn:h}-\ref{eqn:yn}) whereby $|\psi|^2 = |{\cal h}|^2 + \sum_{n=1}^{\cal N} |{\cal y}_n|^2 = \sum_{n=0}^{\cal N}|c_n|^2, $ from which we determine that the gravitational and gauge field wave amplitudes trace a pattern on the surface of an ${\cal N}+1$-dimensional sphere. This behavior is reminiscent of neutrino flavor oscillations, where the mass eigenstates remain in phase while the flavor eigenstates oscillate.

\begin{figure}[ht!]
\center\includegraphics[width=0.435\textwidth,angle=0]{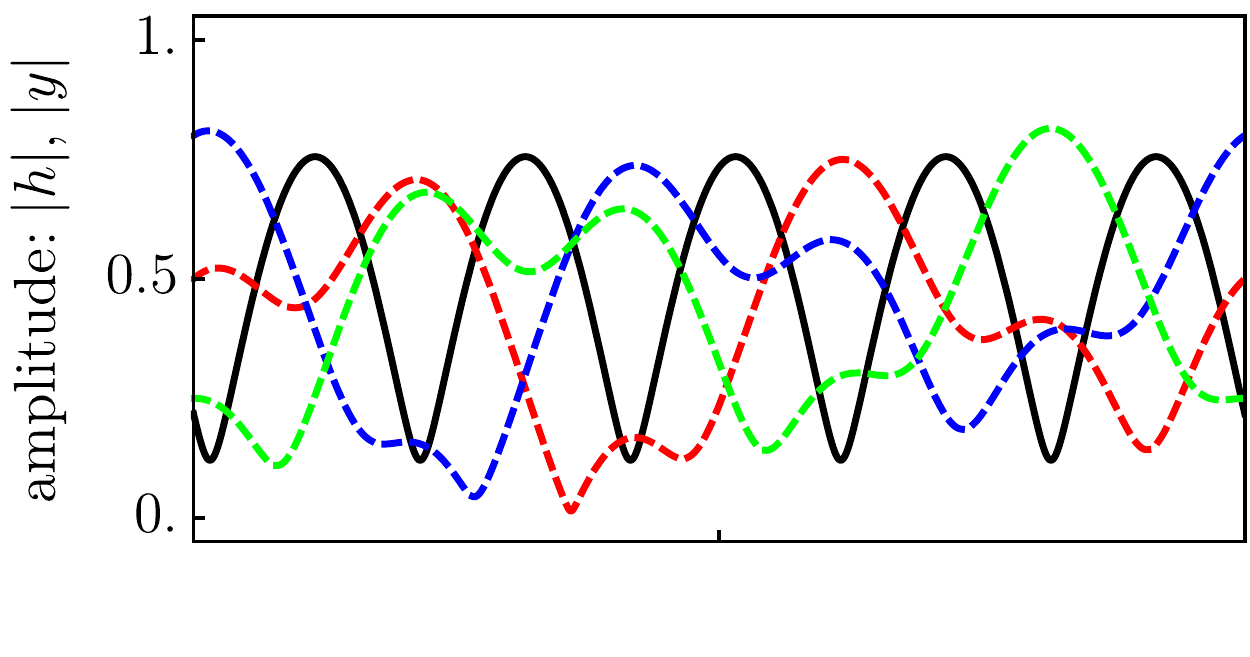}
\vspace{-0.5cm}
\center\includegraphics[width=0.45\textwidth,angle=0]{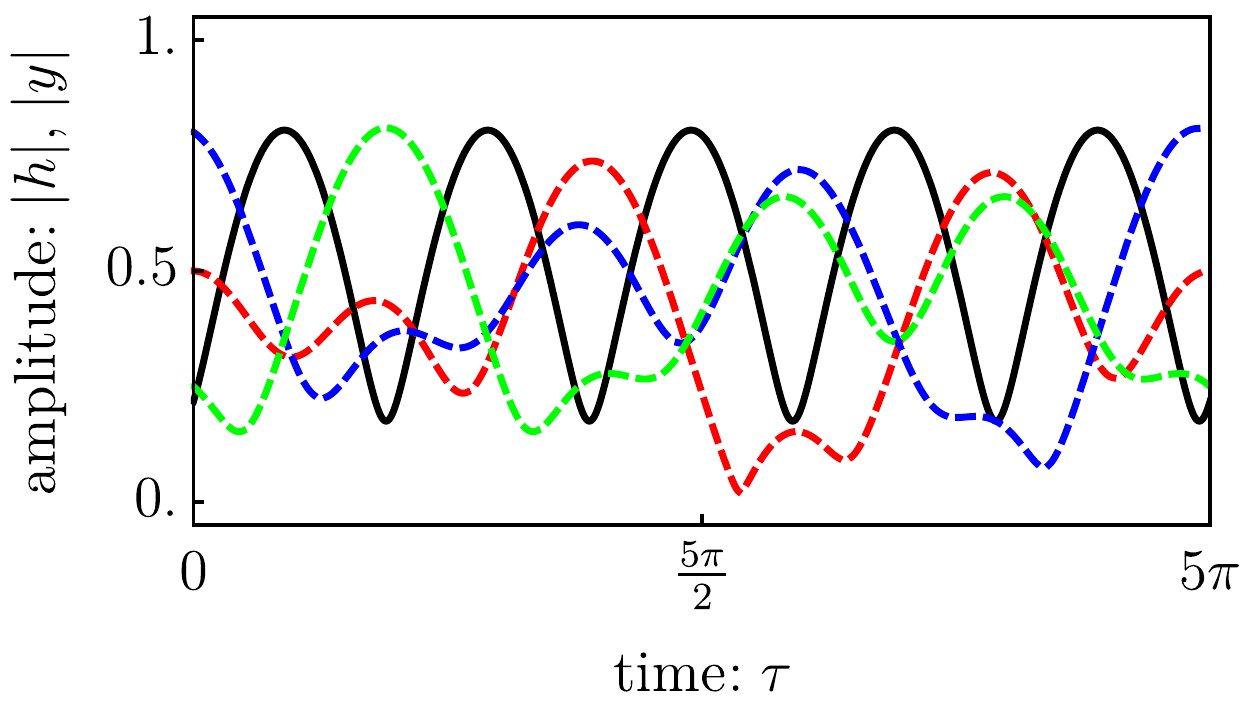}
\caption{The oscillations of the gravitational wave amplitude ${\cal h}$ (black) and gauge field ${\cal y}$ (dashed) are shown for ${\cal N}=3$ and $A_1=A_2=2 |A_3|$. The upper and lower panels show the right- and left-circular polarizations.}
\label{fig:fig2}
\end{figure}

\begin{figure}[hb!]
\center\includegraphics[width=0.45\textwidth,angle=0]{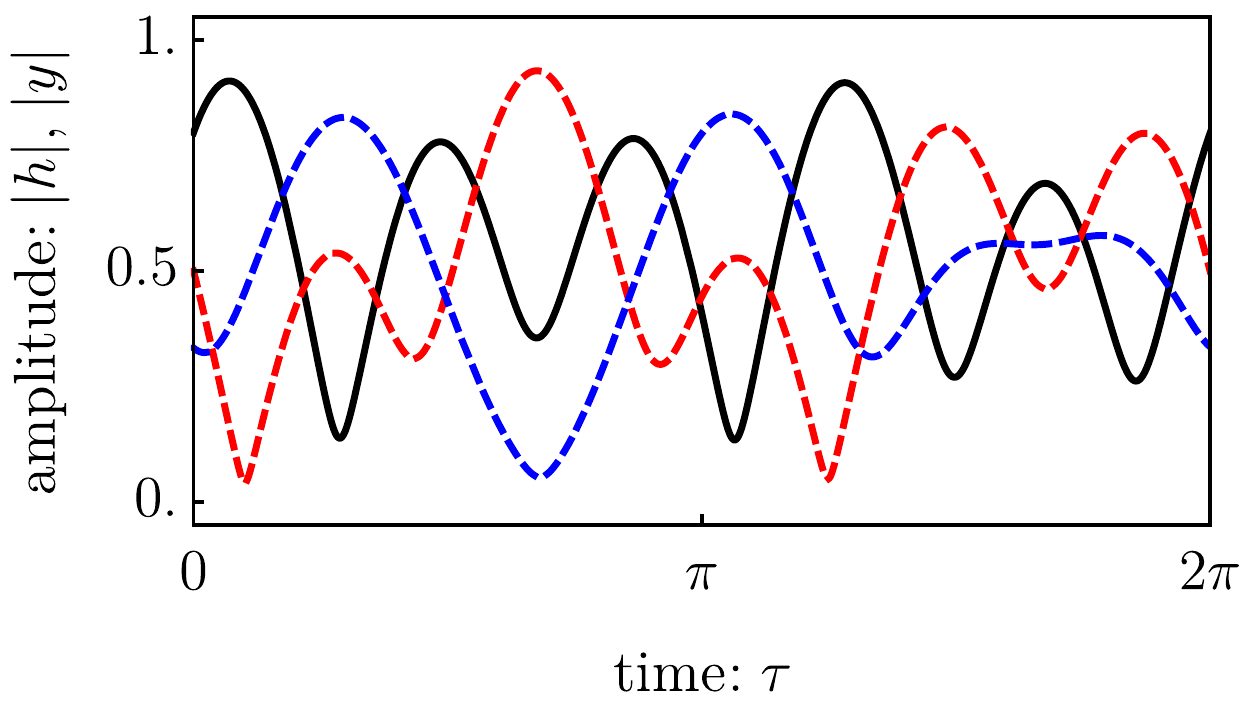}
\caption{The oscillations of the gravitational wave amplitude ${\cal h}$ (black) and gauge field ${\cal y}$ (dashed) are shown for ${\cal N}=2$ with different field strengths and couplings for the two SU(2) subgroups. The gravitational wave responds to these different couplings, and varies between different extrema. The amplitudes are normalized $|\psi|^2=1$ and the coefficients $A_i$ are scaled so that the oscillations repeat with a period $2 \pi$.}
\label{fig:fig3}
\end{figure}

\begin{figure}[ht!]
\begin{frame}{}
\animategraphics[loop,controls,width=0.75\linewidth]{12}{GWGFO60-}{0}{60}
\end{frame} 
\caption{The gravitational wave -- gauge field oscillations in Fig.~2 (upper) are illustrated using oscillating pistons. The gravitational and gauge field waves are represented by the central and surrounding pistons. (View in Adobe Reader to play animation, or see Supplemental Material \cite{supplemental}, or download from Ref.~\cite{download}.)}
\label{fig:fig4}
\end{figure}

We show the time evolution for selected parameters to illustrate the variety of behavior of the gravitational and gauge field amplitudes. In the top panel of Fig.~\ref{fig:fig1}, we show the simplest case, $A_1=1$, $A_2=A_3=0$, as an example of flavor electrodynamics. The second panel shows a second case of flavor electrodynamics, but with two families. In the bottom two panels, we set ${\cal N}=3$ and turn on the coupling $\gy$ such that $A_3 \ll A_2 < A_1$. Between oscillations of the gravitational wave, different gauge field waves dominate. As we introduce $A_3$, the gauge fields oscillate with an additional time scale. In Fig.~\ref{fig:fig2} we show both polarizations for a case with ${\cal N}=3$ and $A_1=A_2=2 |A_3|$. Since $A_3 < \omega$, the gauge fields repeat on a longer time scale. In this specific case, different gauge fields are seen to dominate with seemingly irregular cadence.

Our procedure is easily generalized to cases in which the field strengths and couplings are different for each of the ${\cal N}$ subgroups, although numerical solution of the equations of motion is necessary. Fig.~\ref{fig:fig3} shows such a case for ${\cal N}=2$ with $A_1=A_2=A_3$ for the first SU(2) subgroup (shown in red), and $A_1=4 A_2 = 2 A_3$ for the second (blue). The gravitational wave maintains fixed oscillations, but its height varies between different extrema.
 
To show more than two gauge subgroups, it is useful to visualize the amplitudes directly as oscillators. In Fig.~\ref{fig:fig4} we represent the amplitudes as pistons with height determined by the amplitudes ${\cal h},\, {\cal y}_n$. The placement of the pistons illustrates the relative roles of gravitational and gauge fields. Through this visual tool it is easier to see the interplay between the gauge fields and the pacemaking gravitational wave, and  the implications of a region of stationary gauge fields for gravitational wave physics and cosmology come into focus.

The existence of a new source of tensor modes opens the possibility of a greater variety of gravitational wave spectra in early Universe scenarios. In an inflationary scenario, the procedure for evaluating the primordial gravitational wave spectrum begins by quantizing the high frequency gravitational wave modes (e.g. Ref.~\cite{Baumann:2009ds} for a pedagogic review). In a model involving flavor-space locked gauge fields  \cite{Maleknejad:2011jw,Adshead:2012kp}, a Lagrangian of the form of Eq.~(\ref{eqn:LHY}) describes the gravitational and gauge field waves and their interaction. Quantum fluctuations in the gravitational field give rise to a homogeneous solution $H_{h}$ of the gravitational wave equations of motion, with the amplitude of $c_0$ set by considerations of the Bunch-Davies vacuum, and all other coefficients being zero. Fluctuations in the gauge fields create $n=1,\, ...\,,\,{\cal N}$ inhomogeneous solutions $H_{ih,n}$, each with only one $c_n$ nonzero. Whereas the individual homogeneous and inhomogeneous solutions display the characteristic oscillations that modulate the amplitude, the observable power spectrum is proportional to the sum $|H_h|^2 + \sum_{n=1}^{\cal N} |H_{ih,n}|^2$. The inflationary prescription for quantizing the gravitational and gauge field modes fixes the coefficients $c$ to have equal amplitude, in which case Eqs.~(\ref{eqn:h}-\ref{eqn:yn}) can be used to show the time-dependent modulations cancel out. A different quantization procedure for inflationary fields, and the impact of flavor-space locked gauge field families on early Universe scenarios such as a bouncing cosmology \cite{Battefeld:2014uga} or ekpyrosis \cite{Lehners:2008vx} remain to be explored.

In a scenario in which relic gauge fields contribute a species of dark radiation \cite{Bielefeld:2014nza,Bielefeld:2015daa}, the effect explored in this paper would imprint a novel signature on an inflationary spectrum of primordial gravitational waves. When long wavelength modes enter the horizon and begin to oscillate, a portion of the gravitational wave transforms into a gauge field wave and back again, depending on the relative abundance of the gauge field radiation. Through this process, the modulation of the gravitational wave amplitude in time is translated into a present-day modulation in frequency. Hence, the oscillations seen in Figs.~\ref{fig:fig1}-\ref{fig:fig3} are transferred onto the otherwise power-law gravitational wave spectrum. If all ${\cal N}$ of the gauge field waves are zero at early times, then the behavior is identical to the ${\cal N}=1$ case, illustrated in Fig.~4 of Ref.~\cite{Bielefeld:2015daa}. Not only will the gravitational wave spectrum carry a record of the thermal history of the Universe \cite{Watanabe:2006qe}, it will also carry the imprint of the gravitational wave - gauge field oscillations.

In a cosmological scenario in which dark energy is due to flavor-space locked gauge fields \cite{Mehrabi:2015lfa,Rinaldi:2015iza}, one may expect the dimensionful quantities $A_i$ to be of the order of the comoving Hubble scale. In this case, all subhorizon gravitational waves traveling across cosmological distances are modulated due to the interconversion with gauge fields, including gravitational waves from the binary merger of black holes. Since the gravitational wave amplitude from these astrophysical events have been proposed as a standard siren \cite{Holz:2005df} to determine luminosity distance, our modulation effect leads to a systematic discrepancy with electromagnetic measures of the luminosity distance to an object at the same redshift. In order to estimate the size of this effect, we solve the equations of motion for ${\cal h},\, {\cal y}$ in the case in which the gauge coupling $\gy$ and field strength $\phi$  are dictated by a gauge quintessence scenario \cite{Mehrabi:2015lfa}. We consider a scenario in which the dark energy contributes $\Omega_{DE} \simeq 0.7$ with an average equation of state $\bar w =-0.9$, roughly in line with the constraints in Ref.~\cite{Ade:2015rim}. We express the discrepancy in terms of the luminosity distance, $\Delta h/h = - \Delta d_L/ d_L$, obtaining $\Delta d_L/d_L$ of a few percent for objects out to redshift $z\simeq 1$; a small but non-negligible effect compared to the errors forecast in Ref.~\cite{Holz:2005df}. In this scenario the gauge quintessence also behaves like dark radiation during the radiation-dominated epoch, and so will imprint a frequency modulation on an inflationary spectrum of primordial gravitational waves, as described above. We leave for future work the evaluation of observational constraints and forecast of the impact on standard sirens.

We note that only the gravitational wave affects spacetime geometry and couples universally to other forms of matter. Hence, a gravitational wave detector appropriately positioned would sense a modulated signal, as the gravitational waves convert into gauge fields and back again, for the cases illustrated in Figs.~\ref{fig:fig1}-\ref{fig:fig4}. In principle, fermionic matter charged under the same group as the gauge fields can be used to detect the complementary modulation.

\vfill
\acknowledgments
We thank Peter Adshead for useful conversations.
This work is supported in part by DOE Award No. DE-SC0010386.  
 
\vfill


\end{document}